\documentclass[preprint,12pt]{elsarticle}

\usepackage{graphicx}
\usepackage{amssymb}
\usepackage{amsthm}
\usepackage{mathtools}
\usepackage{amsmath}
\usepackage{lineno}
\usepackage{mathrsfs}
\usepackage{booktabs, caption, makecell}
\usepackage{threeparttable}
\usepackage{algorithm}
\usepackage{algcompatible}
\usepackage{lipsum}
\usepackage{appendix}
\usepackage{float}

\usepackage[nameinlink,capitalise]{cleveref}

\journal{Elsevier}

\begin{document}

\begin{frontmatter}
\title{A Deep Finite Difference Emulator for the Fast Simulation of Coupled Viscous Burgers’ Equation}
\author[label1]{Xihaier Luo\corref{cor1}}
\ead{xluo@bnl.gov}

\author[label1]{Yihui Ren}
\author[label1]{Wei Xu}
\author[label1]{Shinjae Yoo}

\author[label2]{Balasubramanya Nadiga}

\author[label3]{Ahsan Kareem}

\cortext[cor1]{Corresponding author.}

\address[label1]{Computational Science Initiative, Brookhaven National Laboratory, Upton, NY 11973, United States.}
\address[label2]{Los Alamos National Laboratory, Los Alamos, NM 87545, USA.}
\address[label3]{NatHaz Modeling Laboratory, University of Notre Dame, Notre Dame, IN 46556, United States.}

\begin{abstract}
    This work proposes a deep learning-based emulator for the efficient computation of the coupled viscous Burgers' equation with random initial conditions. In a departure from traditional data-driven deep learning approaches, the proposed emulator does not require a classical numerical solver to collect training data. Instead, it makes direct use of the problem's physics. Specifically, the model emulates a second-order finite difference solver, i.e., the Crank–Nicolson scheme in learning dynamics. A systematic case study is conducted to examine the model's prediction performance, generalization ability, and computational efficiency. The computed results are graphically represented and compared to those of state-of-the-art numerical solvers.
\end{abstract}


\end{frontmatter}

\section{Introduction}
\label{sec1}

Burgers’ equation, which consists both of nonlinear wave propagation and diffusive effects, is a leading figure of the most widely used partial differential equations (PDEs) in the modern science and engineering community. With the nonlinear convection term and viscosity term, the coupled Burgers’ equation serves as a viable candidate for analysis in various fields, such as traffic flow, polydispersive sedimentation, turbulence modeling, longitudinal elastic waves in isotropic solids, gas dynamics, and growth of molecular interfaces \cite{cole1951quasi, hirsh1975higher, chowdhury2000statistical, peebles2020large}. However, the nonlinear coupled Burgers’ equation does not possess precise analytic solutions, and developing an accurate numerical approximation is challenging due to nonlinear terms and viscosity parameters. In the past several decades, a wide spectrum of methods for obtaining numerical solutions of the coupled Burgers’ equation has been proposed and tested, including decomposition \cite{kaya2001explicit}, variational iteration \cite{abdou2005variational}, cubic B-spline collocation \cite{mittal2011numerical}, Haar wavelet-based \cite{jiwari2015hybrid}, Chebyshev spectral collocation \cite{khater2008chebyshev}, and fourth-order finite difference methods \cite{bhatt2016fourth}.

Regardless of precision and accuracy, applying a numerical method in practice is steeped in uncertainty arising from a number of sources, such as those due to lack of knowledge, uncertain initial and boundary conditions, or natural material variability \cite{pettersson2009numerical, poette2009uncertainty}. More concretely, we consider the two-dimensional coupled viscous Burgers’ equation

\begin{equation}
\label{Burgers}
\begin{split}
&\frac{\partial u}{\partial t}+u \frac{\partial u}{\partial x}+v \frac{\partial u}{\partial y}= \frac{1}{Re} \left(\frac{\partial^{2} u}{\partial x^{2}}+\frac{\partial^{2} u}{\partial y^{2}}\right), \quad (x, y, t) \in D \times (0, T],\\
&\frac{\partial v}{\partial t}+u \frac{\partial v}{\partial x}+v \frac{\partial v}{\partial y}= \frac{1}{Re} \left(\frac{\partial^{2} v}{\partial x^{2}}+\frac{\partial^{2} v}{\partial y^{2}}\right), \quad (x, y, t) \in D \times (0, T],
\end{split}
\end{equation}

subject to the initial conditions

\begin{equation}
\label{initial_conditions}
\begin{split}
u(x, y, 0)=u_{0}(x, y), \quad (x, y) \in D, \\
v(x, y, 0)=u_{0}(x, y), \quad (x, y) \in D,
\end{split}
\end{equation}

and the boundary conditions

\begin{equation}
\label{boundary_conditions}
\begin{split}
u(x, y, t)=f(x, y, t), \quad (x, y, t) \in \partial D \times(0, T], \\
v(x, y, t)=g(x, y, t), \quad (x, y, t) \in \partial D \times(0, T]
\end{split}
\end{equation}

Randomness is introduced to the initial conditions to better model the target system's stochastic nature, for instance, the presumed chaos in turbulence. Because Fourier series has been applied to approximate the physical turbulence in many studies, we formalize the random initial states using a truncated Fourier series with random coefficients $\boldsymbol{a}, \boldsymbol{b}$, and $\boldsymbol{c} \in \mathbb{R}^2$

\begin{equation}
\label{uncertainty}
\begin{split}
\boldsymbol{w}(x, y) & =\sum_{i=-N_i}^{N_i} \sum_{j=-N_j}^{N_j} \boldsymbol{a}_{i j} \sin (2 \pi(i x+j y))+\boldsymbol{b}_{i j} \cos (2 \pi(i x+j y)) \\
\boldsymbol{u}(x, y, 0) & = \frac{2 \boldsymbol{w}(x, y)}{\max_{\{x, y\}} \boldsymbol{w}(x, y) } + \boldsymbol{c}
\end{split}
\end{equation}

As a result, solving \cref{Burgers} is extended to a case involving repeated queries. To reliably propagate the uncertainty in the input to the quantity of interest, one can use the Monte Carlo (MC) method, which converges in the limit of infinite samples. However, MC approaches require a large number of samples to get convergent statistics, generally in the hundreds of thousands or millions \cite{liu2001monte}. Parallel to the development of more efficient algorithms, a prevalent option is to train an easy-to-evaluate model, namely a surrogate, on some simulation-based data then use the surrogate to conduct prediction and uncertainty propagation tasks instead of solving the actual PDEs \cite{najm2009uncertainty, forrester2009recent}.

In light of this background, deep learning (DL)-based surrogate models have gained tremendous traction, using an ever-growing amount of data to prescribe models. However, a couple of standout issues remain unsolved: (a) training a reliable DL surrogate requires big data, which is not always available for many scientific problems \cite{zhu2019physics, karumuri2020simulator, yang2021b}; (b) most current DL models are either instance-based with limited generalizability, i.e., a new neural network needs to be trained when applied to a new instance \cite{li2020fourier, lu2021learning}, or mesh-dependent, meaning additional interpolation/extrapolation schemes are required to make predictions for points outside of the predefined grids \cite{sirignano2018dgm, esmaeilzadeh2020meshfreeflownet, pfaff2020learning}. To address these issues, we offer a unique autoregressive DL architecture that emulates a chosen numerical solver---in this case, a second-order finite difference approach---to enable mesh-free learning of spatiotemporal dynamics in a data-free manner. Moreover, we assess the proposed model's computational efficiency against conventional numerical solvers on different hardware platforms. A 100-fold improvement in the computational effort is achieved.


\section{Numerical simulation and deep learning emulation}
\label{sec2}
The goal is to create a DL-based surrogate that reduces the need for a numerical solver. This means that the proposed surrogate can learn the coupled viscous Burgers' equation by emulating traditional numerical schemes rather than requiring a large amount of numerical simulation data for training. This section demonstrates the relationship between numerical simulation and DL emulation.

\subsection{Numerical simulation: finite difference method}
\label{sec21}
To construct a numerical solution, consider \cref{Burgers} on a bounded two-dimensional (2D) Cartesian domain $\Omega = [0, L_x] \times [0, L_y]$, where

\begin{equation}
\label{notation}
\begin{split}
& 0 = x_0 < x_1, \dots, x_{n_x-1} < x_{n_x} = L_x, \quad x_{i+1} - x_i = \Delta x,\\
& 0 = y_0 < y_1, \dots, y_{n_y-1} < y_{n_y} = L_y, \, \, \quad y_{j+1} - y_j = \Delta y,\\
& 0 = t_0 < t_1, \dots, t_{n_t-1} < t_{n_t} = T, \quad \quad \, \, \, t_{n+1} - t_n = \Delta t,
\end{split}
\end{equation}

Let nodal points $u_{i, j}^{n}$ and $v_{i, j}^{n}$ be the discrete approximation of $u$ and $v$ at the grid point $(i \Delta x, j \Delta y, n \Delta t)$, respectively. Because the Burgers' equations are time-dependent, a robust and precise numerical approach for temporal discretization is required. Compared to the explicit Euler method, the Crank-Nicolson finite-difference method that allows for the relaxation of strict constraints on the number of time steps and provides more accurate solutions is used \cite{kadalbajoo2006numerical, bahadir2003fully}. By applying Crank–Nicolson scheme to \cref{Burgers}, we get

\begin{equation}
\label{discrete_u}
\begin{split}
\frac{u_{i, j}^{n+1}-u_{i, j}^{n}}{\Delta t} & +\frac{1}{2}\left[u_{i, j}^{n+1}\left(\frac{u_{i+1, j}^{n+1}-u_{i-1, j}^{n+1}}{2 \Delta x}\right)+u_{i, j}^{n}\left(\frac{u_{i+1, j}^{n}-u_{i-1, j}^{n}}{2 \Delta x}\right)\right] \\
& +\frac{1}{2}\left[\operatorname{v}_{i, j}^{n+1}\left(\frac{u_{i, j+1}^{n+1}-u_{i, j-1}^{n+1}}{2 \Delta y}\right)+v_{i, j}^{n}\left(\frac{u_{i, j+1}^{n}-u_{i, j-1}^{n}}{2 \Delta y}\right)\right] \\
&-\frac{1}{\operatorname{Re}}\left[\frac{1}{2}\left\{\left(\frac{u_{i+1, j}^{n+1}-2 u_{i, j}^{n+1}+u_{i-1, j}^{n+1}}{(\Delta x)^{2}}\right)+\left(\frac{u_{i+1, j}^{n}-2 u_{i, j}^{n}+u_{i-1, j}^{n}}{(\Delta x)^{2}}\right)\right\}\right. \\
&+\frac{1}{2}\left\{\left(\frac{u_{i, j+1}^{n+1}-2 u_{i, j}^{n+1}+u_{i, j-1}^{n+1}}{(\Delta y)^{2}}+\left(\frac{u_{i, j+1}^{n}-2 u_{i, j}^{n}+u_{i, j-1}^{n}}{(\Delta y)^{2}}\right)\right\}\right]=0
\end{split}
\end{equation}

and 

\begin{equation}
\label{discrete_v}
\begin{split}
\frac{v_{i, j}^{n+1}-v_{i, j}^{n}}{\Delta t} & +\frac{1}{2}\left[u_{i, j}^{n+1}\left(\frac{v_{i+1, j}^{n+1}-v_{i-1, j}^{n+1}}{2 \Delta x}\right)+u_{i, j}^{n}\left(\frac{v_{i+1, j}^{n}-v_{i-1, j}^{n}}{2 \Delta x}\right)\right] \\
& +\frac{1}{2}\left[\operatorname{v}_{i, j}^{n+1}\left(\frac{v_{i, j+1}^{n+1}-v_{i, j-1}^{n+1}}{2 \Delta y}\right)+v_{i, j}^{n}\left(\frac{v_{i, j+1}^{n}-v_{i, j-1}^{n}}{2 \Delta y}\right)\right] \\
&-\frac{1}{\operatorname{Re}}\left[\frac{1}{2}\left\{\left(\frac{v_{i+1, j}^{n+1}-2 v_{i, j}^{n+1}+v_{i-1, j}^{n+1}}{(\Delta x)^{2}}\right)+\left(\frac{v_{i+1, j}^{n}-2 v_{i, j}^{n}+v_{i-1, j}^{n}}{(\Delta x)^{2}}\right)\right\}\right. \\
&+\frac{1}{2}\left\{\left(\frac{v_{i, j+1}^{n+1}-2 v_{i, j}^{n+1}+v_{i, j-1}^{n+1}}{(\Delta y)^{2}}+\left(\frac{v_{i, j+1}^{n}-2 v_{i, j}^{n}+v_{i, j-1}^{n}}{(\Delta y)^{2}}\right)\right\}\right]=0
\end{split}
\end{equation}

The truncation error of the adopted numerical scheme using the Taylor series expansion is of order $\mathcal{O} ((\Delta x)^2+(\Delta y)^2+(\Delta t)^2)$.

\subsection{Deep learning emulation: convolutions and differential operators}
\label{sec22}

To create a DL model that emulates the mathematical operators in the chosen numerical scheme, we first connect convolution kernels to the discretizations of differential operators. Following the proposition stated in \cite{long2018pde, long2019pde}, consider q to be a filter with sum rules of order $\alpha \in \mathbb{Z}_{+}^{2}$. Then, for a smooth function $F(x)$ on $\mathbb{R}^2$, we have

\begin{equation}
\label{q1}
\frac{1}{\varepsilon^{|\alpha|}} \sum_{k \in \mathbb{Z}^{2}} q[k] F(x+\varepsilon k)=C_{\alpha} \frac{\partial^{\alpha}}{\partial x^{\alpha}} F(x)+O(\varepsilon), \text { as } \varepsilon \rightarrow 0
\end{equation}

In addition, if q has total sum rules of order $K \backslash\{|\alpha|+1\}$ for some $K > |\alpha|$, then

\begin{equation}
\label{q2}
\frac{1}{\varepsilon^{|\alpha|}} \sum_{k \in \mathbb{Z}^{2}} q[k] F(x+\varepsilon k)=C_{\alpha} \frac{\partial^{\alpha}}{\partial x^{\alpha}} F(x)+O\left(\varepsilon^{K-|\alpha|}\right), \text { as } \varepsilon \rightarrow 0
\end{equation}

\cref{q1} and \cref{q2} indicate an $\alpha^{th}$ order differential operator can be approximated by the convolution of a filter with $\alpha^{th}$ order of sum rules. For example, combine \cref{discrete_u} and \cref{discrete_v}, and we have

\begin{equation}
\label{bold}
\mathbf{u}^{t+1} - \mathbf{u}^{t} = \Delta t [ -0.5 f_{x}(\mathbf{u}^{t+1}) + f_{x}(\mathbf{u}^{t}) ]
\end{equation}

with $\mathbf{u} = [u, v]$ and $f_{\Delta x}(\mathbf{u}^{t})$ taking the following form:

\begin{equation}
\label{f_short}
f_{x}(\mathbf{u}^{t}) = \mathbf{u}^{t} (\frac{\partial \mathbf{u}^{t}}{\partial x} + \frac{\partial \mathbf{u}^{t}}{\partial y}) - \frac{1}{Re} (\frac{\partial^{2} \mathbf{u}^{t}}{\partial x^{2}}+\frac{\partial^{2} \mathbf{u}^{t}}{\partial y^{2}})
\end{equation}

With the proposition (\cref{q1} and \cref{q2}) that links the orders of sum rules with the orders of differential operator, the goal is to use convolution filters to obtain an efficiently computable gradient approximation of the differential operators of interest. In our case, the Sobel and Laplace filters can be used to approximate the first- and second-order derivatives in \cref{f_short} with second-order accuracy \cite{canny1986computational, szeliski2010computer}

\begin{equation}
\label{filter}
\begin{split}
\frac{\partial \mathbf{u}}{\partial x} = \frac{1}{8 \Delta x}\left[\begin{array}{lll}-1 & 0 & 1 \\ -2 & 0 & 2 \\ -1 & 0 & 1\end{array}\right] * \mathbf{u} \quad \text{and} \quad \frac{\partial \mathbf{u}}{\partial y} = \frac{1}{8 \Delta y}\left[\begin{array}{lll}-1 & -2 & -1 \\ 0 & 0 & 0 \\ 1 & 2 & 1\end{array}\right] * \mathbf{u} \\
\frac{\partial^2 \mathbf{u}}{\partial x^2} = \frac{1}{2 \Delta x^2}\left[\begin{array}{lll} 1 & 0 & 1 \\ 0 & -4 & 0 \\ 1 & 0 & 1 \end{array}\right] * \mathbf{u} \quad \text{and} \quad \frac{\partial^2 \mathbf{u}}{\partial y^2} = \frac{1}{2 \Delta y^2}\left[\begin{array}{lll} 1 & 0 & 1 \\ 0 & -4 & 0 \\ 1 & 0 & 1 \end{array}\right] * \mathbf{u}
\end{split}
\end{equation}

It is worth noting that filter size determines the computational efficiency and accuracy. Large filters, in general, can approximate differential operators with higher approximation orders. On the other hand, large filters have more memory overhead and a higher computation cost. In practice, the trade-off must be balanced \cite{cai2012image}.

\section{Fourier neural operator}
\label{sec3}
Knowing that convolution kernels can be used to emulate various types of differential operators \cite{long2018pde, long2019pde, cai2012image}, the next step is to create and train a DL model that follows this philosophy. Given a vast collection of neural networks, the emphasis is on computational efficiency and properties similar to a traditional numerical solver. As a result of their lower computational complexity, graph neural networks are preferred over standard artificial neural networks and recurrent neural networks \cite{goodfellow2016deep, kovachki2021neural}. Furthermore, we use a recently developed model, the Fourier neural operator (FNO) \cite{li2020fourier}. Compared to convolutional neural networks, the FNO is mesh-free, like a typical numerical solver. That is, although the model is trained on low-resolution data, it can generate continuous solutions. This section defines and discusses the FNO's architecture and training.

\subsection{Architecture}
\label{sec31}
From a modeling perspective, FNO first employs a local transformation $P$ to lift the state space variable $\mathbf{u}$ to a usually higher-dimensional representation $\mathbf{z}$:

\begin{equation}
\label{lift}
\mathbf{z}_0 = P(\mathbf{u}_{t})
\end{equation}

where $P$ is a shallow fully connected neural network. Like the Koopman operator, the lifting step generalizes the dynamics, facilitating effective learning of the evolution of the state variables. In the case of the Koopman operator, the underlying nonlinear dynamics are assumed to evolve linearly in the lifted space \cite{lusch2018deep, luo2021dynamic}. In this case, one does not impose any restrictions on the lifted evolution. Instead, an iterative refinement is applied to further transform the dynamics $\mathbf{z}_0 \mapsto \mathbf{z}_1 \mapsto \mathbf{z}_2 \mapsto \dots \mapsto \mathbf{z}_N$. The updating rule can be expressed as:

\begin{equation}
\label{iterative}
\mathbf{z}_{j+1}(s):=\sigma\left(W \mathbf{z}_{j}(s)+\left(K_{\phi_{j}} \mathbf{z}_{j}\right)(s)\right), \quad \forall s \in S
\end{equation}

where $\sigma (\cdot)$ is an element-wise $\mathbb{R} \rightarrow \mathbb{R}$ nonlinear activation function, $W (\cdot)$ is a linear transformation, s denotes the coordinates of points of interest in the spatial domain, and $K_{\phi_{j}}$ is a linear operator characterized by a neural network that is parameterized by $\phi_{j} \in \Theta_j$ a finite-dimensional space \cite{battaglia2018relational, wu2020comprehensive}:

\begin{equation}
\label{kernel}
\left(K_{\phi_{j}} \mathbf{z}_{j}\right)(s):=\int_{S} \kappa_{\phi_{J}}(s_1, s_2) \mathbf{z}_{j}(s_2) \mathrm{d} s_2, \quad \forall s \in S
\end{equation}

\begin{figure}[H]
    \centering
    \includegraphics[width=0.9\textwidth]{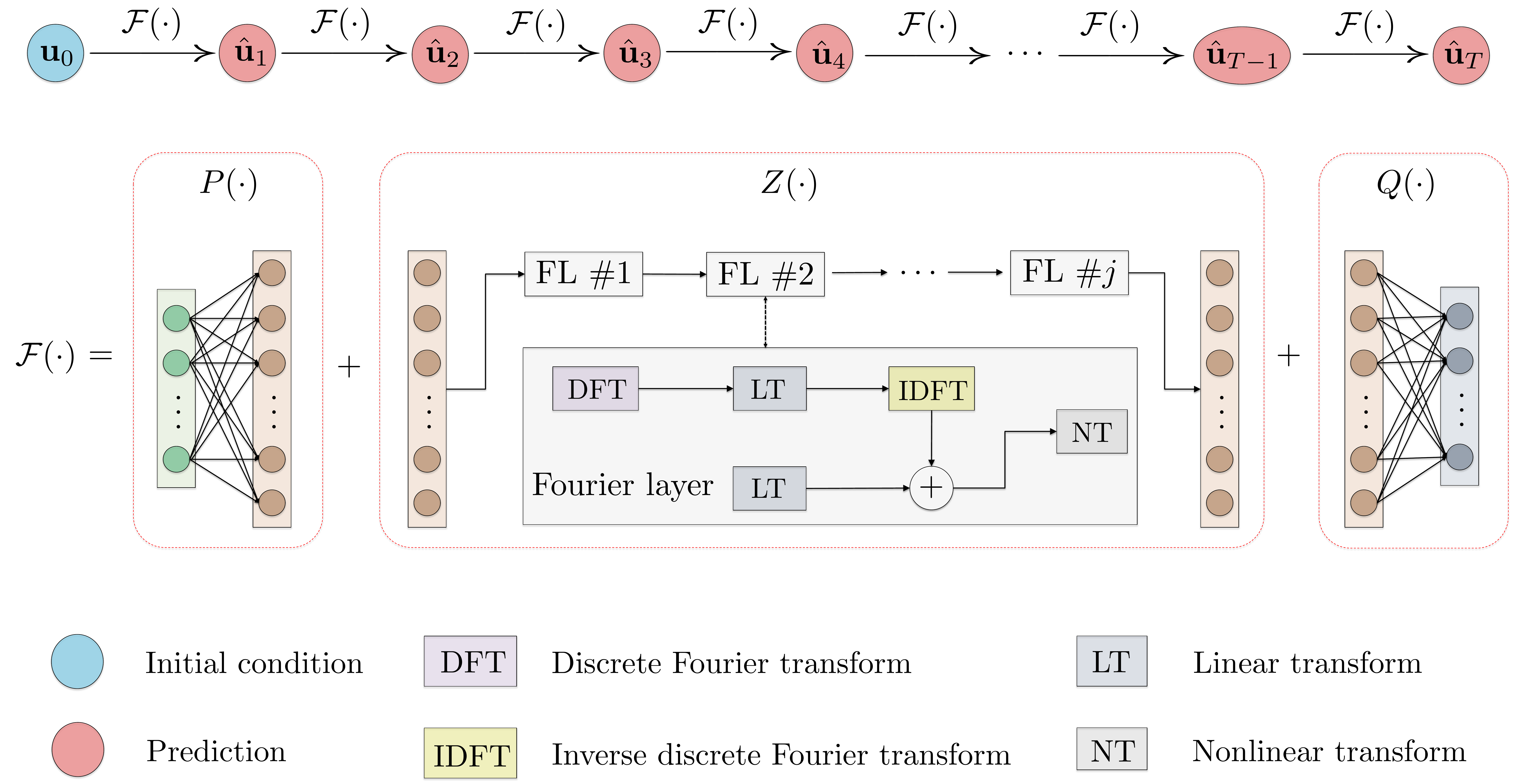}
    \caption{Prediction problem setup and model architecture.}
    \label{fig:problem}
\end{figure}

By letting $\kappa_{\phi_{j}}(s_1, s_2)=\kappa_{\phi_{j}}(s_1-s_2)$ and applying the convolution theorem in Fourier space, Eq. \ref{kernel} becomes \cite{li2020fourier}:

\begin{equation}
\label{FNO}
\left(K_{\phi_{j}}\right)(s)=F^{-1}\left(R_{\phi_{j}} \cdot\left(F \mathbf{z}_{j}\right)\right)(s) \quad \forall s \in S
\end{equation}

with $F(\cdot)$ denoting the fast Fourier transform (FFT) of a function $S \rightarrow \mathbb{R}^{d_\mathbf{z}}$ and $F^{-1}$ representing its inverse calculation. $R_{\phi_{j}}$ is the Fourier transform of a periodic function. This parameterization favors the computation of \cref{kernel}. Lastly, the transformed dynamics are projected back to the state space by another local transformation $Q$:

\begin{equation}
\label{projection}
\mathbf{u}_{t+1} = Q(\mathbf{z}_N)
\end{equation}

where the projection $Q(\cdot)$ again is described by a fully connected neural network. Such operator learning strategy, i.e., learning nonlinear operators mapping one infinite-dimensional Banach space to another, allows for an expressive and efficient architecture that better emulates the underlying dynamics than standard DL methods. \cref{fig:problem} gives a graphic illustration of the emulation problem of
interest and the FNO model described above. For more details regarding the FNO model and operator learning, refer to the work \cite{li2020fourier, kovachki2021neural}.

\subsection{Training}
\label{sec32}
\begin{figure}[H]
    \centering
    \includegraphics[width=0.9\textwidth]{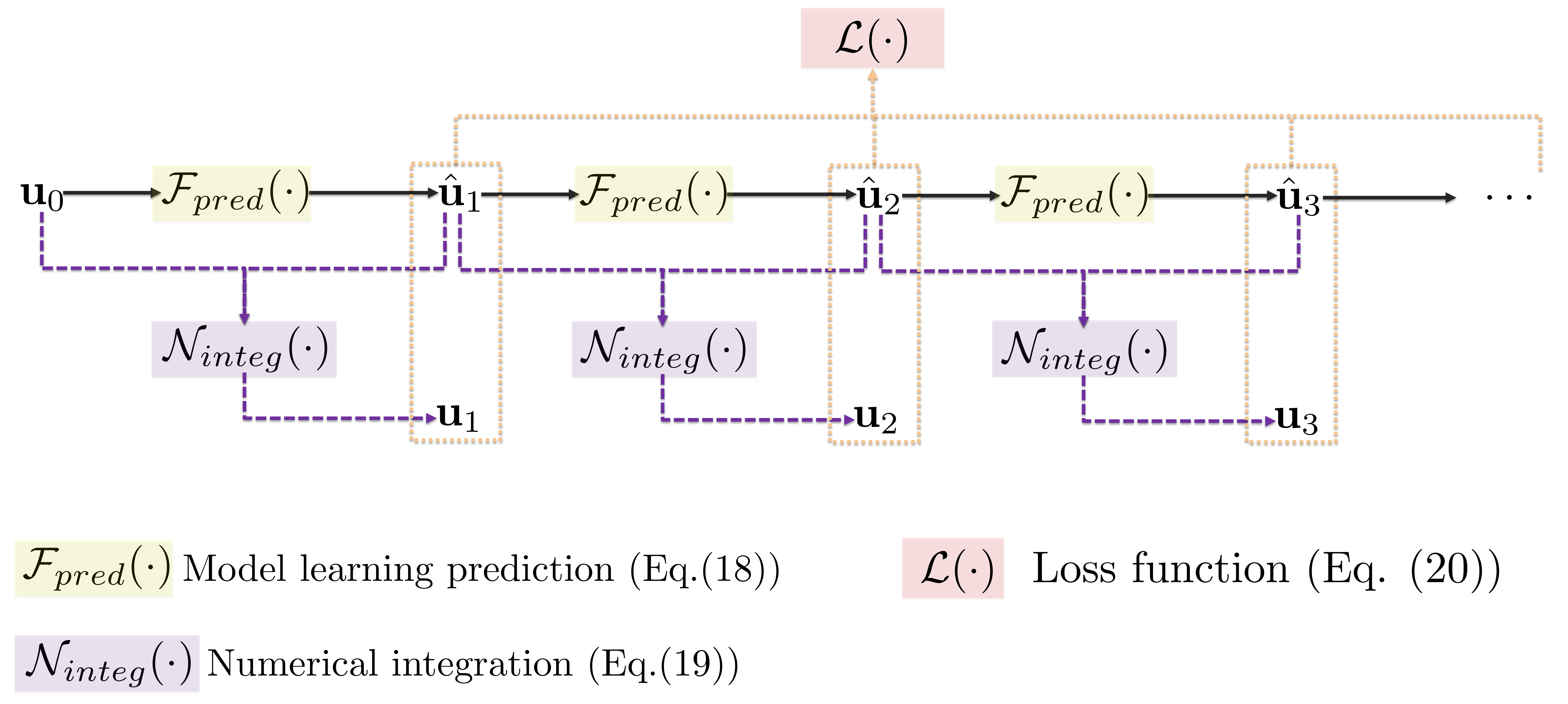}
    \caption{Training of the proposed autoregressive model.}
    \label{fig:training}
\end{figure}

Let $\mathcal{F} (\cdot)$ be an FNO model. Given an initial condition $\mathbf{u}_{0}$, we can implement the model in an autoregressive manner so the prediction of the state variable $\hat{\mathbf{u}}$ at a given time index $t_{k+1}$ can be written as:

\begin{equation}
\label{composition}
\hat{\mathbf{u}}_{k+1} = \underbrace{\left( \mathcal{F} \circ \cdots \circ \mathcal{F} \right)}_{k + 1\text { times }}\left(\mathbf{u}_{0}\right)
\end{equation}

with $\circ$ denoting the composition operation. The goal of training an FNO model is to minimize the difference between the model's prediction and that of a chosen numerical solver. In this work, the Crank–Nicolson method is used for numerical time integration (recall \cref{bold}). For example, we can reformulate \cref{bold} into

\begin{equation}
\label{time_integration}
\mathbf{u}_{k+1} = \mathbf{u}_{k} + \Delta t [ -0.5 f_{x}(\hat{\mathbf{u}}_{k+1}) + f_{x}(\mathbf{u}_{k}) ]
\end{equation}

where $\hat{\mathbf{u}}_{k+1}$ is obtained using \cref{composition}. As previously stated (\cref{f_short} and \cref{filter}), convolutional kernels are used to represent discrete forms of differential operators included in $f_{x}(\cdot)$. For a single time step from $t_{k} \rightarrow t_{k+1}$, we choose to minimize the discretized PDE residual in an $L_2$ norm minimization format \cite{zhu2019physics, geneva2020modeling}. We then arrive at the minimization of the residual of accumulated time steps:

\begin{equation}
\label{physics_constrained}
\mathcal{L}(\boldsymbol{\theta}):=\frac{1}{N_t} \sum_{i=1}^{N_t} ( \mathbf{u}_{i} - \hat{\mathbf{u}}_i )^{2}
\end{equation}

where $\boldsymbol{\theta}$ denotes the learnable model parameters. \cref{fig:training} demonstrates how the loss function is calculated.

\section{Results}
\label{sec4}
This section presents a comprehensive case study to demonstrate the desired properties of the proposed deep finite difference emulator. We first provide the detailed problem setup and model descriptions. Subsequently, we examine the model's prediction performance, generalizability, and computational efficiency. The codes and data used in this work will become available after publication at: https://xihaier.github.io/.

\subsection{Overview}
\label{sec41}
\subsubsection{Problem setup}
\label{sec411}
Consider the coupled viscous Burgers’ equation (\cref{Burgers}) on a bounded 2D Cartesian domain $\Omega = [0, 1] \times [0, 1]$ with boundary conditions

\begin{equation}
\label{bcs}
\mathbf{u} (x, 0, t) = \mathbf{u} (x, 1, t), \quad \mathbf{u} (0, y, t) = \mathbf{u}(1, y, t), \quad t \in [0, 1]
\end{equation}

and initial conditions given in \cref{initial_conditions}. The Reynolds number is kept constant, $Re = 200$, across experiments, and randomness is introduced into the initial conditions via a truncated Fourier series (\cref{uncertainty}), where random coefficients are defined as

\begin{equation}
\label{random_coefficients}
\boldsymbol{a}_{i j}, \boldsymbol{b}_{i j} \sim \mathcal{N}\left(0, \boldsymbol{I}_{2}\right), \quad \boldsymbol{c} \sim \mathcal{U}(-1,1) \in \mathbb{R}^{2}, \quad N_i = N_j = 4
\end{equation}

Several numerical simulations with different grid sizes $64 \times 64$, $128 \times 128$, $256 \times 256$, and $512 \times 512$ are performed to test the spatial accuracy and computational speedup order of the proposed emulator. Unless otherwise specified, the default numerical solver simulates the system until $t = 1$ on a $64 \times 64$ 2D grid with a time step of $0.005$.

\begin{figure}[b!]
    \centering
    \includegraphics[width=0.7\textwidth]{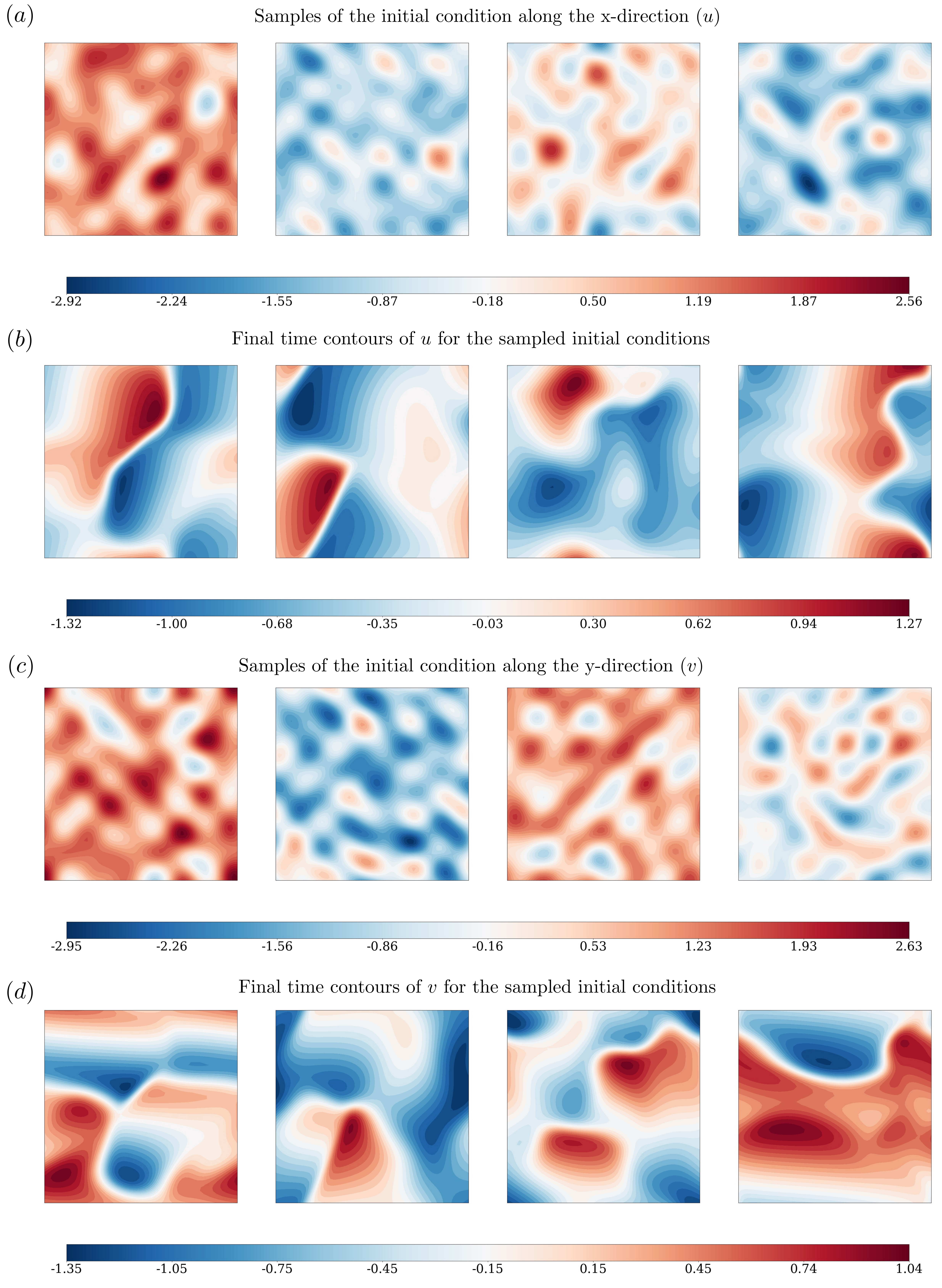}
    \caption{Four random initial conditions and their corresponding final time velocity fields.}
    \label{fig:initials}
\end{figure}

We plot four simulation results to show how the initial uncertainty affects the dynamics. \cref{fig:initials} (a) and (c) display four random samples of initial conditions of $u$ and $v$ generated by \cref{uncertainty}. Note that these initial numbers and fluid structures vary across samples. More importantly, the system evolves uniquely in terms of waveforms and fluid motions. \cref{fig:initials} (b) and (d) give the contours representing the final time velocity field for the selected initial conditions. The different velocity fields affirm the challenge of capturing stochastic advection effects on long-term dynamics \cite{cole1951quasi}.

\subsubsection{Baseline}
\label{sec412}
We compare the adopted FNO model to the state-of-the-art model, ResNet, to demonstrate the desirable properties. After extensive hyperparameter optimization, we configure the two models as follows:

\begin{itemize}
    \item \textit{ResNet}: A 27-layer residual convolutional neural network \cite{he2016deep}. The final dense layer is defined as a convolutional layer with two output channels, representing predictions of $u$ and $v$, respectively. 
    \item \textit{FNO}: A four-layer Fourier neural operator with two fully connected networks, (FCN) $P (\cdot)$ (\cref{lift}) and $Q (\cdot)$ (\cref{projection}) \cite{li2020fourier}. The first FCN lifts the dynamics from $\mathbb{R}^{d_{in}}$ to $\mathbb{R}^{20}$, and the last FCN projects the dynamics from  $\mathbb{R}^{20}$ to $\mathbb{R}^{2}$. Also, each Fourier layer is parameterized with $12$ truncated modes for the matrix-vector multiplication involved in the computation of $R_{\phi_{j}}$ (\cref{FNO}). 
\end{itemize}

We generate $N_{train} = 8192$ training instances and another $N_{test} = 512$ testing instances for all experiments, where an instance denotes an initial condition. Note each experiment includes two regions, namely, the training $[0, 0.5]$ and the extrapolation $(0.5, 1]$. That is, $N_t$ in \cref{physics_constrained} is 100, and the trained model is used to predict the next 100 time steps. Overall, the dynamics are projected 200 steps in time by the emulator. An Adam variant called \textit{decoupled weight decay regularization} \cite{loshchilov2018decoupled} is adopted as the optimizer with a weight decay $\lambda = 0.0001$. Both models are trained for $120$ epochs with an initial learning rate of $0.001$ that decays by $\gamma = 0.995$ every epoch.

\subsection{Prediction performance}
\label{sec42}

\subsubsection{Qualitative evaluation}
\label{sec421}

\begin{figure}[H]
    \centering
    \includegraphics[width=0.95\textwidth]{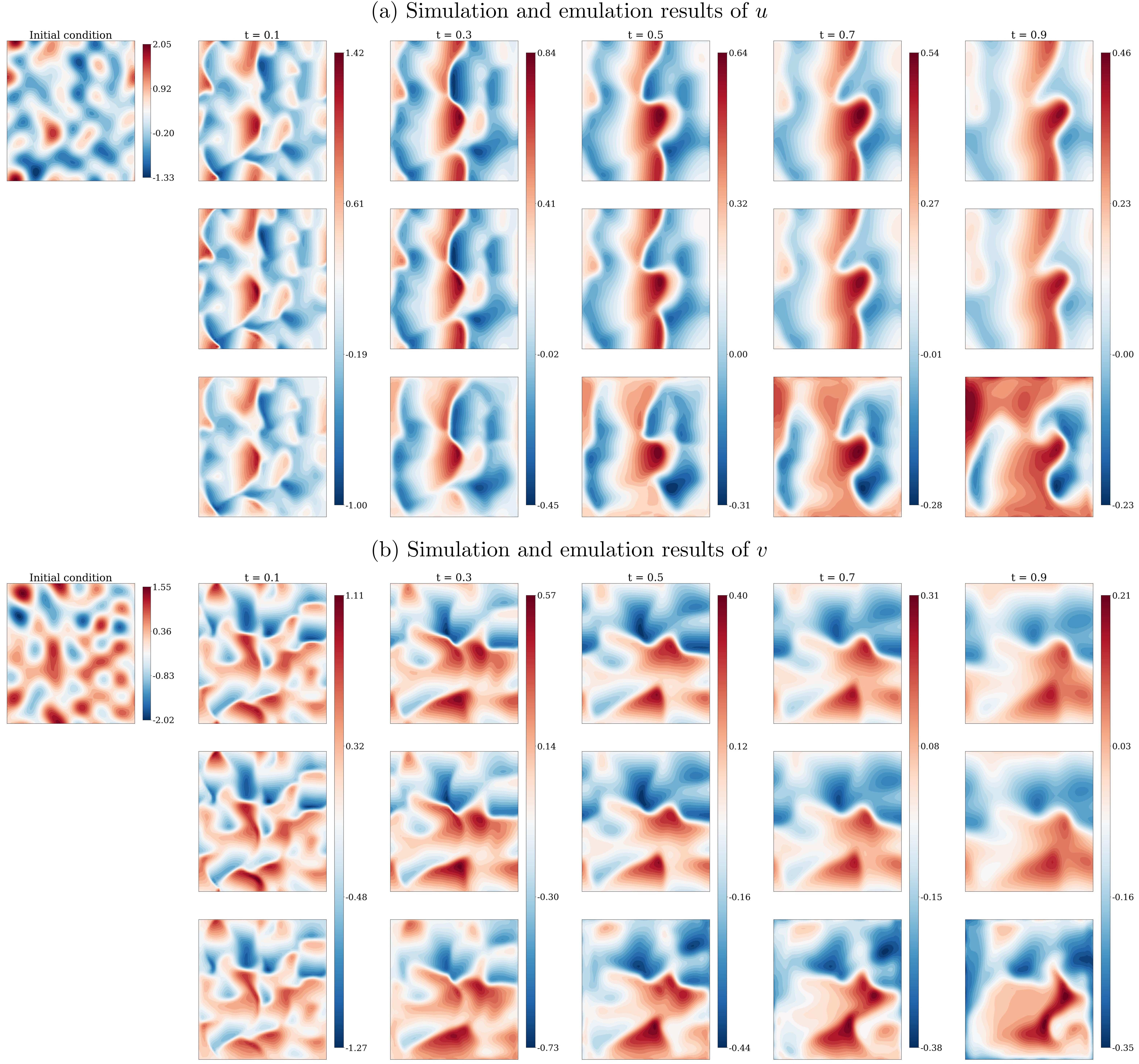}
    \caption{Four random initial conditions and their corresponding final time velocity fields.}
    \label{fig:prediction}
\end{figure}

\cref{fig:prediction} shows the emulation results of a randomly chosen realization from the test dataset. Obviously, the adopted FNO model outperforms the state-of-the-art ResNet model trained on the same amount of initial condition realizations. In \cref{fig:prediction}, it can be observed that both models are able to provide satisfactory predictions in the predefined training time period. However, ResNet's performance degrades significantly in the extrapolation region. ResNet's predictions deviate from the ground truth, while the prediction results of FNO still can maintain relatively high accuracy. These results reflect FNO's superior ability to absorb underlying physical principles, which supports better generalization of unknown scenarios \cite{li2020fourier, pfaff2021learning, sanchez2020learning}.

\subsubsection{Quantitative evaluation}
\label{sec422}
To quantitatively test the present scheme's precision, the root-mean-square errors (RMSEs) and relative errors (REs) are reported. They are defined as

\begin{equation}
\label{RMSE}
\operatorname{RMSE}=\sqrt{\sum_{x} \sum_{y} \big( \mathbf{u} (x,y,t) - \hat{\mathbf{u}} (x,y,t) \big)}
\end{equation}

and 

\begin{equation}
\label{metREric}
\operatorname{RE} = \bigg| \frac{ \sum_{x} \sum_{y} \big( \mathbf{u} (x,y,t) - \hat{\mathbf{u}} (x,y,t) \big) }{\sum_{x} \sum_{y} \mathbf{u}(x,y,t) } \bigg|
\end{equation}

We tested $512$ different samples of the fluid realizations. \cref{tab:mean} and \cref{tab:std} show the mean and standard deviation of the aforementioned two error metrics. It is observed that FNO outperforms ResNet by an order of magnitude when it comes to controlling error propagation. Although both models' accuracy decreases at a similar rate in the extrapolation region, the FNO model has a significantly lower mean value of RMSE/RE, particularly in the training region. Furthermore, the FNO model has a lower standard deviation than ResNet. This suggests that the FNO model is more reliable and robust in emulating Burgers' equations.

\begin{table}[H]
    \centering
    \includegraphics[width=0.95\textwidth]{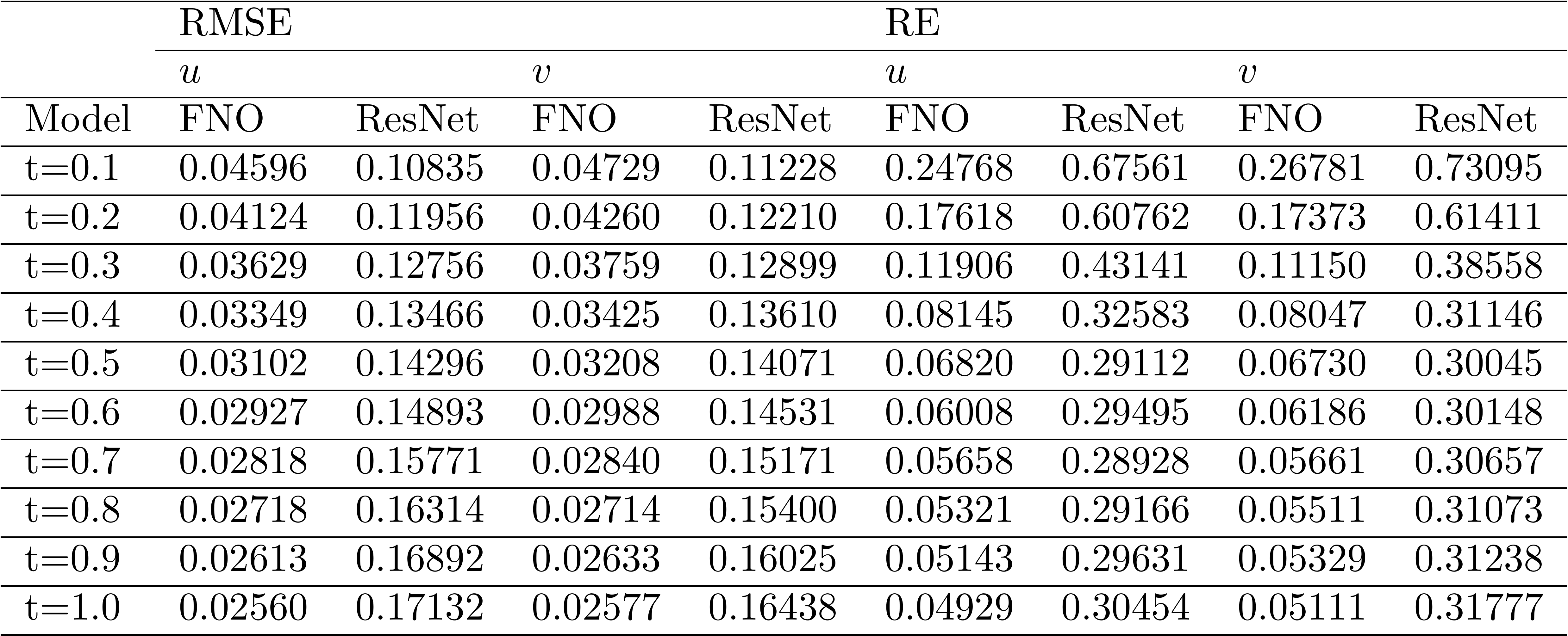}
    \caption{The mean value of root-mean-square errors and relative errors for $u$ and $v$ at different time instances.}
    \label{tab:mean}
\end{table}

\begin{table}[H]
    \centering
    \includegraphics[width=0.95\textwidth]{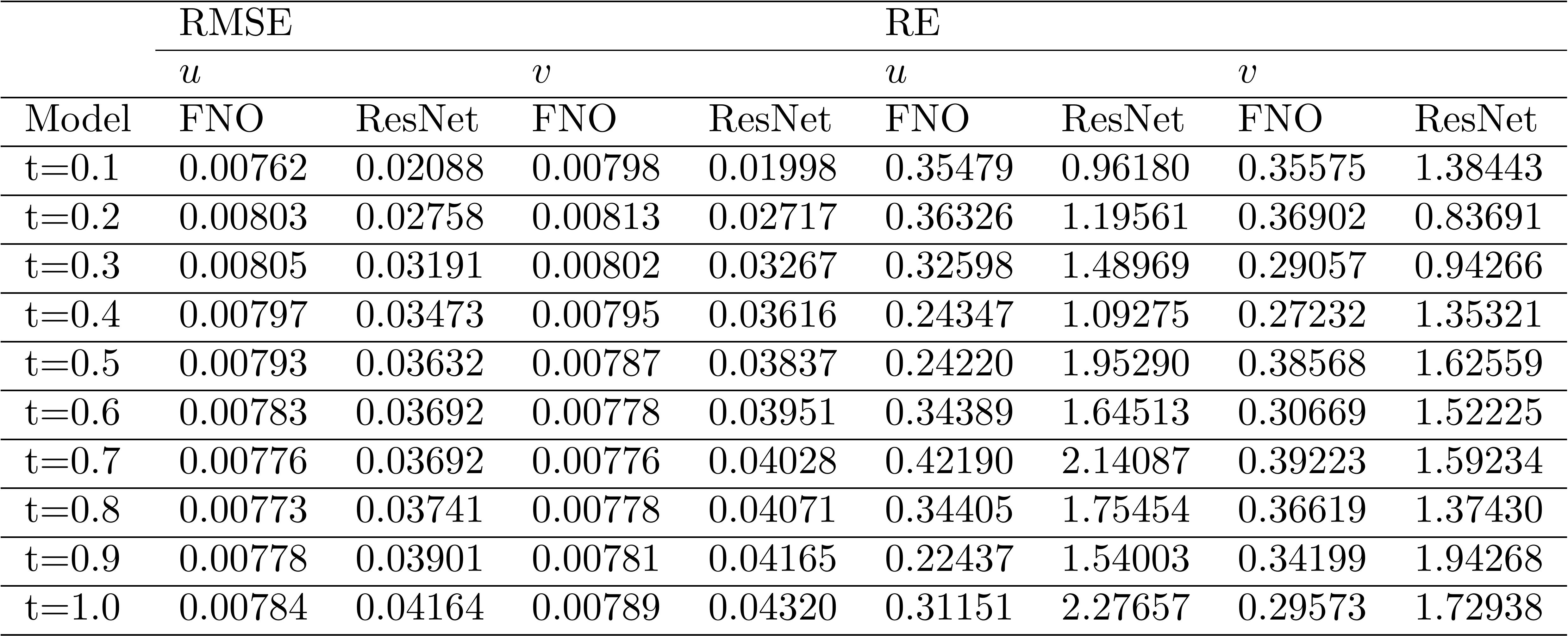}
    \caption{The standard deviation of root-mean-square errors and relative errors for $u$ and $v$ at different time instances.}
    \label{tab:std}
\end{table}

\subsection{Generalization performance}
\label{sec43}

\subsubsection{Out-of-distribution prediction}
\label{sec431}

Here, we investigate the emulators' generalization performance to \textit{out-of-distribution} initial conditions. We take the emulators trained on random inputs defined in \cref{random_coefficients} and test them on new input distributions:

\begin{equation}
\label{new_coefficients}
\boldsymbol{a}_{i j}, \boldsymbol{b}_{i j} \sim \mathcal{U}(0, 1) \in \mathbb{R}^{2}, \quad \boldsymbol{c} \sim \mathcal{U}(-1,1) \in \mathbb{R}^{2}, \quad N_i = N_j = 4
\end{equation}

\begin{figure}[H]
    \centering
    \includegraphics[width=0.97\textwidth]{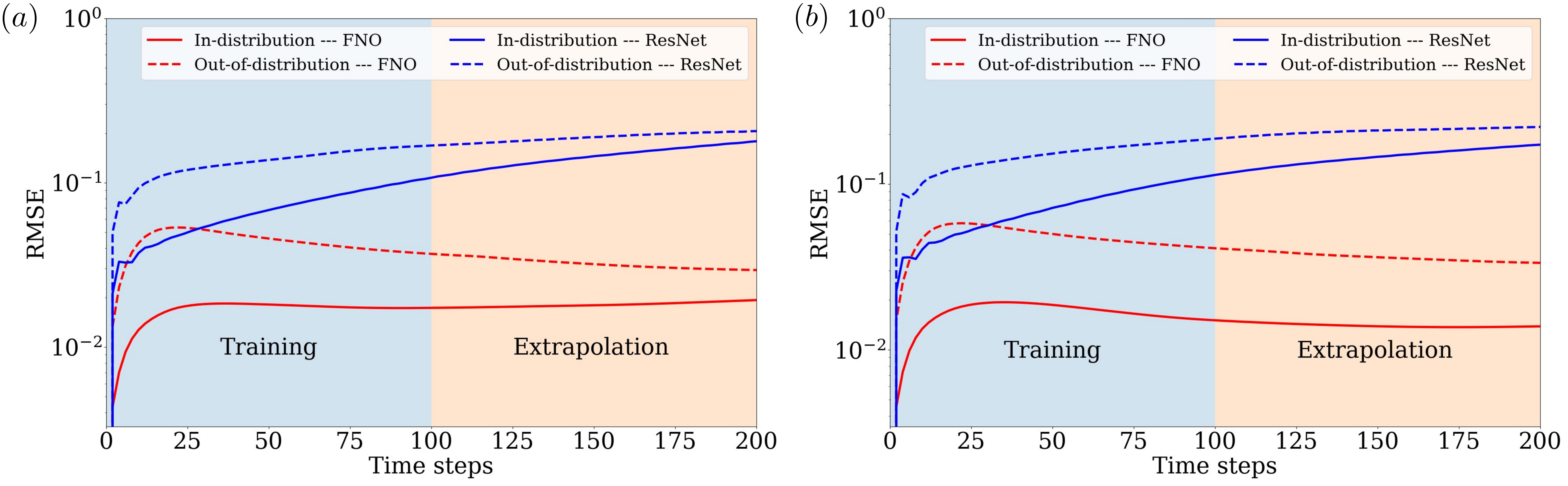}
    \caption{The mean of root-mean-square errors for (a) $u$ and (b) $v$.}
    \label{fig:out_of_dis}
\end{figure}

\cref{fig:out_of_dis} shows the mean RMSE of the emulation results of $u$ and $v$. While both emulators are affected by \textit{out-of-distribution} samples, the FNO model has lower errors. Notably, it is observed that FNO outperforms the ResNet in controlling the error propagation approximately by order of magnitude of 10. This effect gets more pronounced in the extrapolation region, where the mean RMSEs of the ResNet model increase as the system evolves. This finding is consistent with the intuitive expectation of using a surrogate model to forecast nonlinear dynamics. However, FNO is able to stabilize, even reduce, the prediction error within the same time region. In the meantime, it should be noted that FNO \textit{out-of-distribution} prediction performance outperforms ResNet in-distribution test performance. These findings suggest that the FNO model possesses superior generalization performance than ResNet.

\begin{figure}[H]
    \centering
    \includegraphics[width=0.95\textwidth]{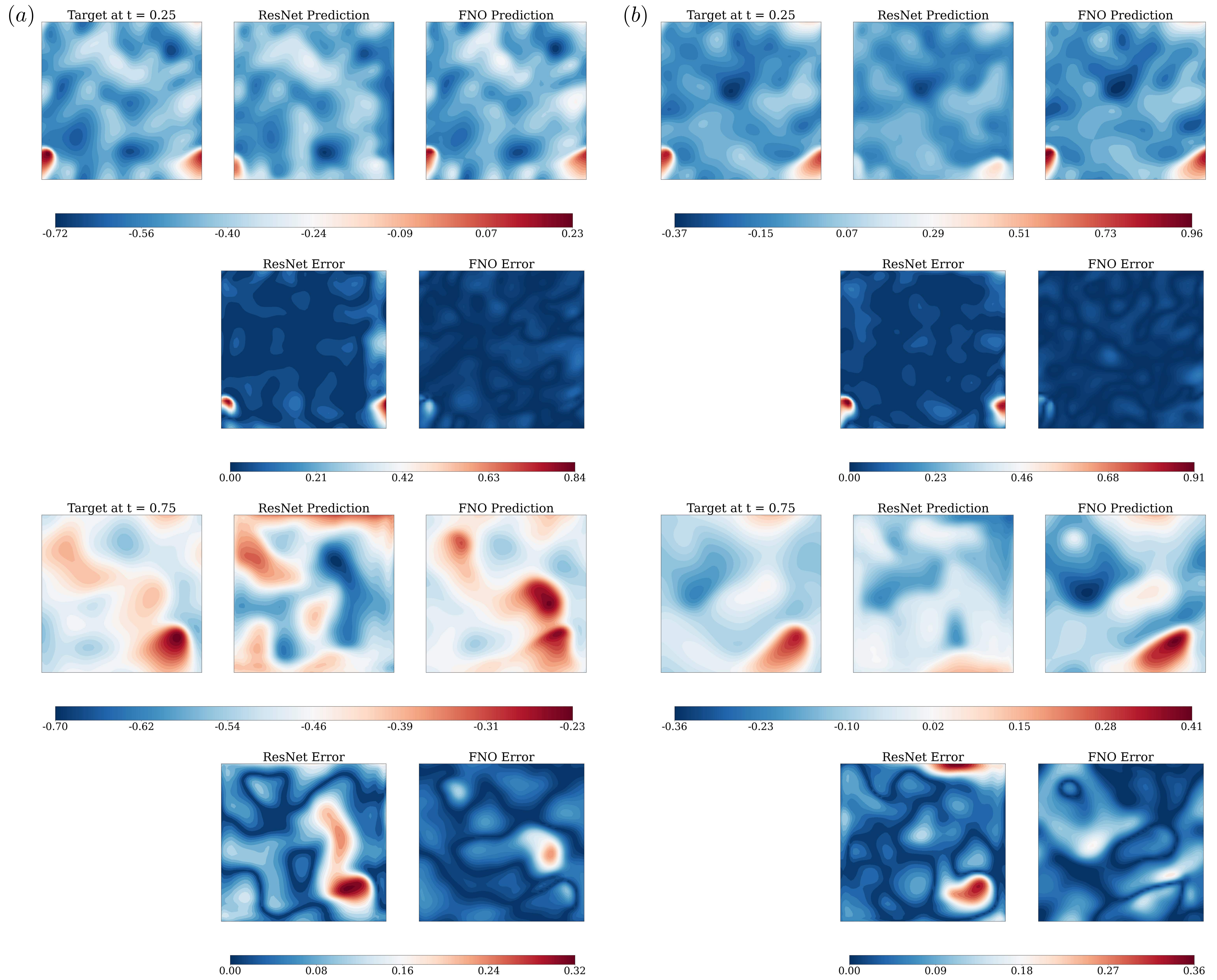}
    \caption{Predictions of the (a) x-velocity and (b) y-velocity of the selected \textit{out-of-distribution} sample.}
    \label{fig:generalization2}
\end{figure}

\cref{fig:generalization2} additionally shows the prediction performance for a randomly chosen realization with time index specified to $0.25$ for training and $0.75$ for extrapolation. The FNO predictions are almost unaffected by the \textit{out-of-distribution} samples, while the ResNet model seriously deteriorates and produces worse predictions, especially in the extrapolation region. The ResNet model's error contour also shows the bulk of errors is clustered on the leading face of the waves, specifically where the extreme values are located \cite{geneva2020modeling}. However, the errors of the FNO predictions are relatively small and evenly distributed, and there is no clear high-error area in the predictions. These findings indicate that the FNO generalizes better to unknown initial conditions than ResNet.

\subsubsection{Super-resolution prediction}
\label{sec432}
\begin{figure}[H]
    \centering
    \includegraphics[width=0.93\textwidth]{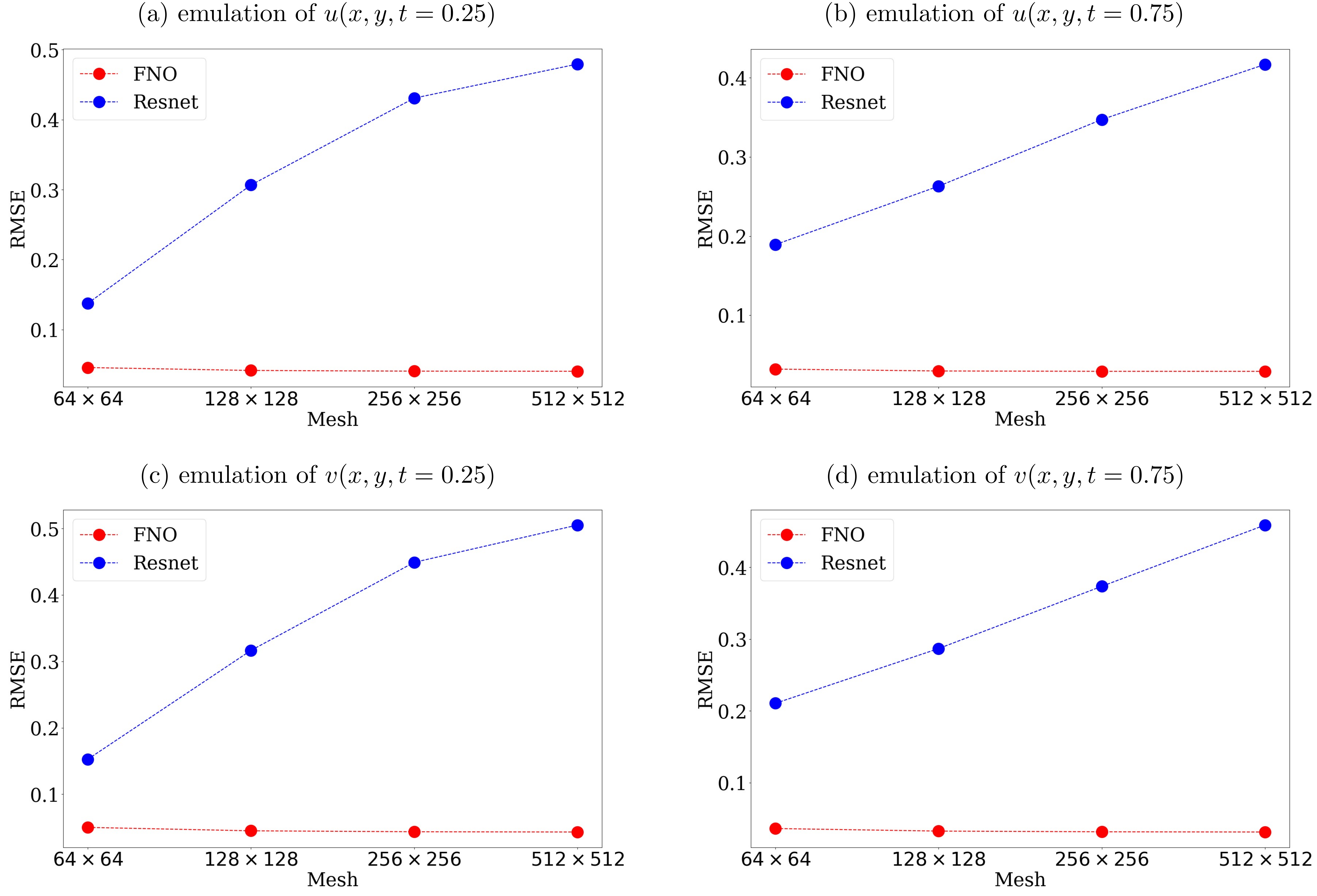}
    \caption{Scaling performance on discretization from $64 \times 64$ up to $512 \times 512$.}
    \label{fig:super_resolution}
\end{figure}

This section summarizes the performance of the two previously described emulators on super-resolution prediction tasks \cite{esmaeilzadeh2020meshfreeflownet}. The goal is to evaluate the generalization ability of these two emulators on various discretization meshes. ResNet and FNO are both trained on $64 \times 64$ meshes and tested at higher resolutions. \cref{fig:super_resolution} depicts the results. Of note, the error is defined as the pixel-wise error averaged across all spatial grids of a given time instance. We chose $t=0.25$ as the training region's representative and $t=0.75$ as the extrapolation region's representative. The emulation results show that because convolution kernels are spatial-agnostic and channel-specific, ResNet fails to provide accurate prediction if additional tuning for different resolutions is not available \cite{goodfellow2016deep}. When evaluated at a higher resolution, the proposed FNO model, on the other hand, achieves consistent error. The promising results show that the FNO has greater generalization potential as the model is tailored to learn operators mapping between infinite-dimensional function spaces rather than standard finite-dimensional Euclidean spaces \cite{sanchez2020learning}.

\subsection{Computational efficiency}
\label{sec44}
\begin{figure}[H]
    \centering
    \includegraphics[width=0.75\textwidth]{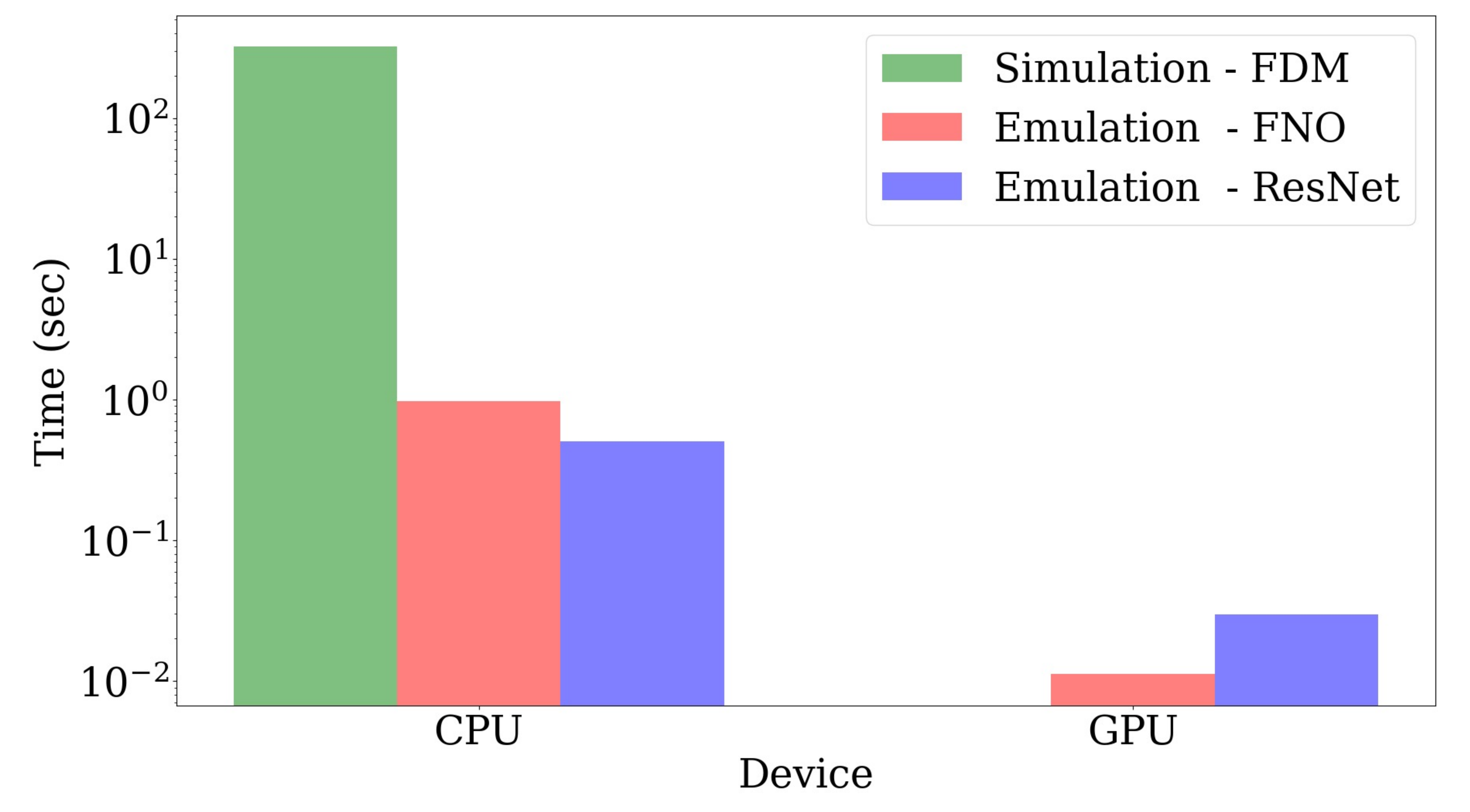}
    \caption{Computation time for different experiments. The time refers to the wall clock time for inferring one sample.}
    \label{fig:computation_time}
\end{figure}

As previously mentioned, advances in DL have benefited from rapid development in both the central processing unit (CPU) and graphics processing unit (GPU), which provides powerful computing resources and speeds up calculations. In this section, a study of the proposed model's simulation efficiency is carried out. Two computation platforms are scheduled: (1) an Intel Xeon Processor E5-2698 v3 with a system memory of 187 GB and (2) a NVIDIA A100 GPU with 40 GB HBM2 memory \footnote{https://www.nersc.gov/systems/perlmutter/}. The total computational cost is classified into two broad categories: training and testing. First, the computational cost of DL-based emulators is compared against the standard numerical simulation. The computation time for a single simulation/emulation of the 2D coupled Burgers' equation on a $64 \times 64$ spatial gridded domain is shown in \cref{fig:computation_time}. Utilizing the same CPU, the speedup of the FNO model is approximately more than 100 times compared to the numerical solver for a single run. More significant saving on the computational effort is achieved if inferences are made using the GPU device. A speedup of three to four orders of magnitude is accomplished. These encouraging results show the potential of using a DL-based emulator for problems involving massive queries of the simulation model.

\section{Conclusions}
\label{sec5}
In this work, we propose a novel physics-constrained learning algorithm for emulating the system of 2D Burgers’ equations. Combining the state-of-the-art Fourier neural operator with classical numerical schemes, we can fast obtain numerical solutions. Furthermore, the proposed emulator is mesh-invariant, allowing the model to be trained on a coarse resolution yet deployed to a fine resolution without sacrificing accuracy. More importantly, the proposed loss function is derived from variational principles. There is no need to collect simulation data via a typical numerical solver. Instead, we cast the Burgers’ equations into a discretized form. The explicit goal is to minimize the difference between the model's prediction and a pre-chosen numerical time integration method. This is implicitly analogous to the $L_2$ minimization of the discretized PDE residual. The numerical case study shows the proposed emulator performed well in various testing scenarios. One obvious future direction for this model is to extend it to three dimensions, as well as other complex physical systems. It would be worthwhile to investigate parallel computation to improve the algorithm's efficiency. Simultaneously, development of more robust graphical convolution and differentiation algorithms also may broaden the scope of the proposed model.

\section*{Acknowledgments}
The authors are grateful for the professional assistance received from Charity Plata in the editing of this paper. This work is supported by the U.S. Department of Energy (DOE), Office of Science, Advanced Scientific Computing Research under Award Number DE-SC-0012704. Brookhaven National Laboratory is supported by the DOE’s Office of Science under Contract No. DE-SC0012704. This research used Perlmutter supercomputer of the National Energy Research Scientific Computing Center, a DOE Office of Science User Facility supported by the Office of Science of the U.S. Department of Energy under Contract No. DE-AC02-05CH11231 using NERSC award NERSC DDR-ERCAP0022110.

\bibliographystyle{elsarticle-num}
\bibliography{sample}

\begin{thebibliography}{10}
\expandafter\ifx\csname url\endcsname\relax
  \def\url#1{\texttt{#1}}\fi
\expandafter\ifx\csname urlprefix\endcsname\relax\def\urlprefix{URL }\fi
\expandafter\ifx\csname href\endcsname\relax
  \def\href#1#2{#2} \def\path#1{#1}\fi

\bibitem{cole1951quasi}
J.~D. Cole, On a quasi-linear parabolic equation occurring in aerodynamics,
  Quarterly of applied mathematics 9~(3) (1951) 225--236.

\bibitem{hirsh1975higher}
R.~S. Hirsh, Higher order accurate difference solutions of fluid mechanics
  problems by a compact differencing technique, Journal of computational
  physics 19~(1) (1975) 90--109.

\bibitem{chowdhury2000statistical}
D.~Chowdhury, L.~Santen, A.~Schadschneider, Statistical physics of vehicular
  traffic and some related systems, Physics Reports 329~(4-6) (2000) 199--329.

\bibitem{peebles2020large}
P.~J.~E. Peebles, The large-scale structure of the universe, Vol.~98, Princeton
  university press, 2020.

\bibitem{kaya2001explicit}
D.~Kaya, An explicit solution of coupled viscous burgers' equation by the
  decomposition method, International Journal of Mathematics and Mathematical
  Sciences 27~(11) (2001) 675--680.

\bibitem{abdou2005variational}
M.~Abdou, A.~Soliman, Variational iteration method for solving burger's and
  coupled burger's equations, Journal of computational and Applied Mathematics
  181~(2) (2005) 245--251.

\bibitem{mittal2011numerical}
R.~Mittal, G.~Arora, Numerical solution of the coupled viscous burgers’
  equation, Communications in Nonlinear Science and Numerical Simulation 16~(3)
  (2011) 1304--1313.

\bibitem{jiwari2015hybrid}
R.~Jiwari, A hybrid numerical scheme for the numerical solution of the
  burgers’ equation, Computer Physics Communications 188 (2015) 59--67.

\bibitem{khater2008chebyshev}
A.~Khater, R.~Temsah, M.~Hassan, A chebyshev spectral collocation method for
  solving burgers’-type equations, Journal of Computational and Applied
  Mathematics 222~(2) (2008) 333--350.

\bibitem{bhatt2016fourth}
H.~P. Bhatt, A.-Q.~M. Khaliq, Fourth-order compact schemes for the numerical
  simulation of coupled burgers’ equation, Computer Physics Communications
  200 (2016) 117--138.

\bibitem{pettersson2009numerical}
P.~Pettersson, G.~Iaccarino, J.~Nordstr{\"o}m, Numerical analysis of the
  burgers’ equation in the presence of uncertainty, Journal of Computational
  Physics 228~(22) (2009) 8394--8412.

\bibitem{poette2009uncertainty}
G.~Po{\"e}tte, B.~Despr{\'e}s, D.~Lucor, Uncertainty quantification for systems
  of conservation laws, Journal of Computational Physics 228~(7) (2009)
  2443--2467.

\bibitem{liu2001monte}
J.~S. Liu, J.~S. Liu, Monte Carlo strategies in scientific computing, Vol.~10,
  Springer, 2001.

\bibitem{najm2009uncertainty}
H.~N. Najm, Uncertainty quantification and polynomial chaos techniques in
  computational fluid dynamics, Annual review of fluid mechanics 41 (2009)
  35--52.

\bibitem{forrester2009recent}
A.~I. Forrester, A.~J. Keane, Recent advances in surrogate-based optimization,
  Progress in aerospace sciences 45~(1-3) (2009) 50--79.

\bibitem{zhu2019physics}
Y.~Zhu, N.~Zabaras, P.-S. Koutsourelakis, P.~Perdikaris, Physics-constrained
  deep learning for high-dimensional surrogate modeling and uncertainty
  quantification without labeled data, Journal of Computational Physics 394
  (2019) 56--81.

\bibitem{karumuri2020simulator}
S.~Karumuri, R.~Tripathy, I.~Bilionis, J.~Panchal, Simulator-free solution of
  high-dimensional stochastic elliptic partial differential equations using
  deep neural networks, Journal of Computational Physics 404 (2020) 109120.

\bibitem{yang2021b}
L.~Yang, X.~Meng, G.~E. Karniadakis, B-pinns: Bayesian physics-informed neural
  networks for forward and inverse pde problems with noisy data, Journal of
  Computational Physics 425 (2021) 109913.

\bibitem{li2020fourier}
Z.~Li, N.~B. Kovachki, K.~Azizzadenesheli, K.~Bhattacharya, A.~Stuart,
  A.~Anandkumar, et~al., Fourier neural operator for parametric partial
  differential equations, in: International Conference on Learning
  Representations, 2020.

\bibitem{lu2021learning}
L.~Lu, P.~Jin, G.~Pang, Z.~Zhang, G.~E. Karniadakis, Learning nonlinear
  operators via deeponet based on the universal approximation theorem of
  operators, Nature Machine Intelligence 3~(3) (2021) 218--229.

\bibitem{sirignano2018dgm}
J.~Sirignano, K.~Spiliopoulos, Dgm: A deep learning algorithm for solving
  partial differential equations, Journal of computational physics 375 (2018)
  1339--1364.

\bibitem{esmaeilzadeh2020meshfreeflownet}
S.~Esmaeilzadeh, K.~Azizzadenesheli, K.~Kashinath, M.~Mustafa, H.~A. Tchelepi,
  P.~Marcus, M.~Prabhat, A.~Anandkumar, et~al., Meshfreeflownet: a
  physics-constrained deep continuous space-time super-resolution framework,
  in: SC20: International Conference for High Performance Computing,
  Networking, Storage and Analysis, IEEE, 2020, pp. 1--15.

\bibitem{pfaff2020learning}
T.~Pfaff, M.~Fortunato, A.~Sanchez-Gonzalez, P.~Battaglia, Learning mesh-based
  simulation with graph networks, in: International Conference on Learning
  Representations, 2020.

\bibitem{kadalbajoo2006numerical}
M.~K. Kadalbajoo, A.~Awasthi, A numerical method based on crank-nicolson scheme
  for burgers’ equation, Applied mathematics and computation 182~(2) (2006)
  1430--1442.

\bibitem{bahadir2003fully}
A.~R. Bahad{\i}r, A fully implicit finite-difference scheme for two-dimensional
  burgers’ equations, Applied Mathematics and Computation 137~(1) (2003)
  131--137.

\bibitem{long2018pde}
Z.~Long, Y.~Lu, X.~Ma, B.~Dong, Pde-net: Learning pdes from data, in:
  International Conference on Machine Learning, PMLR, 2018, pp. 3208--3216.

\bibitem{long2019pde}
Z.~Long, Y.~Lu, B.~Dong, Pde-net 2.0: Learning pdes from data with a
  numeric-symbolic hybrid deep network, Journal of Computational Physics 399
  (2019) 108925.

\bibitem{canny1986computational}
J.~Canny, A computational approach to edge detection, IEEE Transactions on
  pattern analysis and machine intelligence~(6) (1986) 679--698.

\bibitem{szeliski2010computer}
R.~Szeliski, Computer vision: algorithms and applications, Springer Science \&
  Business Media, 2010.

\bibitem{cai2012image}
J.-F. Cai, B.~Dong, S.~Osher, Z.~Shen, Image restoration: total variation,
  wavelet frames, and beyond, Journal of the American Mathematical Society
  25~(4) (2012) 1033--1089.

\bibitem{goodfellow2016deep}
I.~Goodfellow, Y.~Bengio, A.~Courville, Deep learning, MIT press, 2016.

\bibitem{kovachki2021neural}
N.~Kovachki, Z.~Li, B.~Liu, K.~Azizzadenesheli, K.~Bhattacharya, A.~Stuart,
  A.~Anandkumar, Neural operator: Learning maps between function spaces, arXiv
  preprint arXiv:2108.08481 (2021).

\bibitem{lusch2018deep}
B.~Lusch, J.~N. Kutz, S.~L. Brunton, Deep learning for universal linear
  embeddings of nonlinear dynamics, Nature communications 9~(1) (2018) 1--10.

\bibitem{luo2021dynamic}
X.~Luo, A.~Kareem, Dynamic mode decomposition of random pressure fields over
  bluff bodies, Journal of Engineering Mechanics 147~(4) (2021) 04021007.

\bibitem{battaglia2018relational}
P.~W. Battaglia, J.~B. Hamrick, V.~Bapst, A.~Sanchez-Gonzalez, V.~Zambaldi,
  M.~Malinowski, A.~Tacchetti, D.~Raposo, A.~Santoro, R.~Faulkner, et~al.,
  Relational inductive biases, deep learning, and graph networks, arXiv
  preprint arXiv:1806.01261 (2018).

\bibitem{wu2020comprehensive}
Z.~Wu, S.~Pan, F.~Chen, G.~Long, C.~Zhang, S.~Y. Philip, A comprehensive survey
  on graph neural networks, IEEE transactions on neural networks and learning
  systems 32~(1) (2020) 4--24.

\bibitem{geneva2020modeling}
N.~Geneva, N.~Zabaras, Modeling the dynamics of pde systems with
  physics-constrained deep auto-regressive networks, Journal of Computational
  Physics 403 (2020) 109056.

\bibitem{he2016deep}
K.~He, X.~Zhang, S.~Ren, J.~Sun, Deep residual learning for image recognition,
  in: Proceedings of the IEEE conference on computer vision and pattern
  recognition, 2016, pp. 770--778.

\bibitem{loshchilov2018decoupled}
I.~Loshchilov, F.~Hutter, Decoupled weight decay regularization, in:
  International Conference on Learning Representations, 2019.

\bibitem{pfaff2021learning}
T.~Pfaff, M.~Fortunato, A.~Sanchez-Gonzalez, P.~Battaglia, Learning mesh-based
  simulation with graph networks, in: International Conference on Learning
  Representations, 2021.

\bibitem{sanchez2020learning}
A.~Sanchez-Gonzalez, J.~Godwin, T.~Pfaff, R.~Ying, J.~Leskovec, P.~Battaglia,
  Learning to simulate complex physics with graph networks, in: International
  Conference on Machine Learning, PMLR, 2020, pp. 8459--8468.

\end{thebibliography}

\end{document}